\documentclass[11pt,a4paper,leqno]{amsart}
\usepackage{calrsfs}
\usepackage{stmaryrd}
\usepackage{bm}
\usepackage{graphicx}
\usepackage{xcolor}

\swapnumbers
\numberwithin{equation}{section}

\newcommand{\cb}[1]{\color{blue}#1\color{black}}
\renewcommand{\cb}[1]{#1}

\newcommand{\p}{\partial}
\newcommand{\N}{\mathbb{N}}
\newcommand{\R}{\mathbb{R}}
\renewcommand{\d}{\,\mathrm{d}}

\newcommand{\eps}{\varepsilon}
\renewcommand{\epsilon}{\varepsilon}
\newcommand{\la}{\left\langle}
\newcommand{\ra}{\right\rangle}

\newcommand{\Q}{\mathbf{Q}}
\renewcommand{\P}{\mathbf{P}}
\newcommand{\E}{\mathbf{E}}
\newcommand{\Cov}{\mathbf{Cov}}
\newcommand{\1}{\mathbf{1}}
\newcommand{\F}{\mathcal{F}}
\newcommand{\T}{\mathcal{T}}

\newcommand{\A}{\mathcal{A}}
\newcommand{\si}{\mathrm{si}}
\newcommand{\nds}{\mathrm{nds}}
\newcommand{\as}{\mathrm{as}}
\newcommand{\de}{\mathrm{de}}

\providecommand{\newoperator}[3]{\newcommand*{#1}{\mathop{#2}#3}}
\newoperator{\esssup}{\mathrm{esssup}}{\limits}
\DeclareMathOperator*{\wick}{\diamond}

\begin{document}


\theoremstyle{definition}
\newtheorem{dfn}[equation]{Definition}
\theoremstyle{plain}
\newtheorem{thm}[equation]{Theorem}
\newtheorem{pro}[equation]{Proposition}
\newtheorem{cor}[equation]{Corollary}
\newtheorem{lmm}[equation]{Lemma}
\theoremstyle{remark}
\newtheorem{rem}[equation]{Remark}
\newtheorem{exa}[equation]{Example}


\setlength{\abovedisplayskip}{1ex plus .2ex minus .2ex}
\setlength{\belowdisplayskip}{1ex plus .2ex minus .2ex}
\setlength{\jot}{0.5ex}

 
\title{Fractional processes as  models in stochastic finance}

\author[Bender, C.]{Christian Bender}
\address{Christian Bender\\
Department of Mathematics \\ 
Saarland University\\
P.O.Box 151150\\ 
66041 Saarbr\"{u}cken\\ 
Germany}
\email{bender@math.uni-sb.de}

\author[Sottinen, T.]{Tommi Sottinen}
\address{Tommi Sottinen\\
Department of Mathematics and Statistics\\ 
University of Vaasa
P.O.Box 700\\ 
65101 Vaasa\\ 
Finland}
\email{tommi.sottinen@uwasa.fi}

\author[Valkeila, E.]{Esko Valkeila}
\address{Esko Valkeila\\
Department of Mathematics and Systems Analysis\\ 
Aalto University\\ 
P.O.Box 11100\\
00076 Aalto \\
Finland}
\email{esko.valkeila@tkk.fi}

\begin{abstract}
We survey some new progress on the pricing models driven by fractional Brownian motion \cb{or }mixed fractional Brownian motion. In particular, we give results on arbitrage
opportunities, hedging, and option pricing in these models. We summarize some recent results on fractional Black \& Scholes pricing model with transaction costs. We end the paper by giving some approximation results and indicating some open problems related to the paper.  
\end{abstract}
\thanks{T.S. and E.V. acknowledge the support from Saarland University, and E.V. is grateful to the
Academy of Finland, grant 127634, for partial support.}
\maketitle

{\small
\noindent\textbf{JEL Classification:} G10, G13

\noindent\textbf{Mathematics Subject Classification (2000):} 91B28, 91B70, 60G15, 60H05

\noindent\textbf{Keywords}:Fractional Brownian motion, arbitrage, hedging in fractional models, approximation of geometric fractional Brownian motion

}


 
\section{Introduction}

The classical Black-Scholes pricing model is based on standard geometric Brownian motion.
The log-returns of this model are independent and Gaussian. Various empirical studies on the statistical properties of log-returns show that the log-returns are not necessarily independent and also not Gaussian. One way to a more realistic modelling is to change the geometric Brownian motion to a geometric fractional Brownian motion: the dependence of the log-return increments can now be modelled with the Hurst parameter of the fractional Brownian motion.  But then the  pricing model admits arbitrage possibilities with continuous trading, and also with certain discrete type trading strategies. 

The arbitrage possibilities with continuous trading depend on the notion of stochastic integration theory used in the definition of trading strategy. If these stochastic integrals are interpreted as Skorohod integrals, then the arbitrage possibilities with continuous trading disappear. We will not consider this approach in what follows. For a summary of the results obtained in this area we refer to two recent monographs on fractional Brownian motion 
\cite{BiaginiHuOksendalZhang08} and \cite{Mishura08}. If one uses Skorohod integration theory, then one has several problems with the financial interpretation of these continuous trading strategies. We refer to the above two monographs for more details on these issues; see also \cite{BjorkHult05} and \cite{SottinenValkeila03} for the critical remarks concerning the Skorohod approach from the finance point of view.

In this work we discuss the arbitrage possibilities in the fractional Black-Scholes pricing model and in the related mixed Brownian--fractional Brownian pricing model. Then we consider hedging of options in these models.
The fractional Black-Scholes model admits strong arbitrage, and this implies that the initial wealth for the exact hedging strategy cannot be interpreted as a price of the option. But these replication results are interesting from the mathematical point of view. With proportional transaction costs the arbitrage possibilities disappear in the fractional Black-Scholes pricing model.  Hence it is of some interest to know the hedging strategy without transaction costs. For the mixed Brownian-fractional Brownian pricing models the arbitrage possibilities are not that obvious, and the hedging price can be sometimes interpreted as the price of the option.
We shall review some recent results related to these questions. 

One possibility to study the properties of the fractional Black-Scholes pricing model is to approximate it with simpler pricing models. We will present some results on the approximation at the end of this work.  

\section{Models and notions of arbitrage}

\begin{dfn}\label{dfn:fBm}
The \emph{fractional Brownian motion} (fBm) with \emph{Hurst index} 
$H\in(0,1)$ is the centered Gaussian process $B=(B_t)_{t\in[0,T]}$
with $B_0=0$ and 
$$
\Cov\left[B_t , B_s\right] =
\frac12\left(t^{2H}+s^{2H}-|t-s|^{2H}\right).
$$
\end{dfn}

\begin{rem}\label{rem:fBm-properties}
Some well-known properties of the fBm are:
\begin{enumerate}
\item
The fBm has stationary increments.
\item 
For $H=1/2$ the fBm is the standard \emph{Brownian motion} (Bm) $W$.
\item
If $H\ne 1/2$ the fBm is not a semimartingale (cf.
\cite[Theorem 3.2]{Cheridito03} or Example \ref{exa:fBS-arbitrage-3}
later).
\item
If $H>1/2$ the fBm has zero \emph{quadratic variation} (QV) 
(cf. Definition \ref{dfn:QV} later). If $H<1/2$ the QV is $+\infty$.
For the Bm case $H=1/2$ the QV is identity.
\item
For $H>1/2$ the fBm has \emph{long range dependence} (LRD) in the sense 
that
$$
\rho(n) = \Cov\left[B_k-B_{k-1},B_{k+n}-B_{k+n-1}\right]
$$
satisfies
$$
\sum_{n=1}^\infty \left|\rho(n)\right| = +\infty.
$$
\item
The paths of the fBm are a.s. H\"older continuous with index $H-\eps$,
where $H$ is the Hurst index and $\eps$ is any positive constant,
but not H\"older continuous with index $H$. The first claim
follows from the Kolmogorov--Chentsov criterion, and the second
claim follows from the law of iterated logarithm of \cite{Arcones95}:
$$
\limsup_{t\downarrow 0} \frac{B_t}{t^H\sqrt{2\ln \ln 1/t}}=1
\quad\mbox{a.s.}.
$$
\item
The fBm is \emph{self-similar} with index $H$, i.e. for all $a>0$,
$$
\mathrm{Law}\left( (a^H B_{at})_{t\in [0,T/a]}\right)
= \mathrm{Law}\left( (B_t)_{t\in[0,T]}\right).
$$ 
Actually, the fBm is the (up to a multiplicative constant) unique centered
Gaussian self-similar process with stationary increments.
\end{enumerate}
\end{rem}

In this survey we shall consider the following three discounted 
stock-price models in parallel:

\begin{dfn}\label{dfn:models}
Let $S=(S_t)_{t\in[0,T]}$ be the discounted stock-price.
\begin{enumerate}
\item
In the \emph{Black--Scholes model} (BS model)
$$
S_t = s_0e^{\mu t + \sigma W_t - \frac12\sigma^2t},
$$
where $W$ is a Bm, and $s_0,\sigma>0$, $\mu\in\R$.
\item
In the \emph{fractional Black--Scholes model} (fBS model)
$$
S_t = s_0e^{\mu t + \nu B_t},
$$
where $B$ is a fBm with $H\ne 1/2$, and $s_0,\nu>0$, $\mu\in\R$.
\item
In the \emph{mixed fractional Black--Scholes model} (mfBS model)
$$
S_t = s_0e^{\mu t + \sigma W_t -\frac12\sigma^2 t + \nu B_t},
$$
where $W$ is a Bm, $B$ is a fBm with $H\ne 1/2$, $W$ and $B$ are
independent, and $s_0,\sigma,\nu>0$, $\mu\in\R$.
\end{enumerate}
\end{dfn}

\begin{rem}\label{rem:simplicity}
We shall often, for the sake of simplicity and without loss of any
real generality, assume that $\mu=0$ and $\sigma=\nu=s_0=1$.
\end{rem}

\begin{rem}\label{rem:mfBS-vs-fBS}
\begin{enumerate}
\item
The mfBS model is similar to the fBS model in the sense that they have 
essentially the same covariance structure. So, in particular, if $H>1/2$, 
they both have LRD characterized by the Hurst index $H$. 
\item
The fBS model and the mfBS are different in the sense that the mfBS model 
has the same QV as the BS model (cf. Proposition \ref{pro:QV-mfBS}) when
$H > 1/2$. But 
the fBS model has $0$ QV for $H>1/2$.
So, while the fBS model and the mfBS model have the same statistical
LRD property, the pricing in these models is different, in the fBS model it might even be impossible.
\end{enumerate}
\end{rem}

We shall work, except in Section \ref{sect:appr}, in the canonical 
stochastic basis  $(\Omega,\F,(\F_t)_{t\in[0,T]},\P)$. So, 
$\Omega=C_{s_0}^+([0,T])$ the space of positive continuous functions over
$[0,T]$ starting from $s_0$, and the stock-price is the co-ordinate 
process: $S_t(\omega)=\omega_t$. The filtration $(\F_t)_{t\in[0,T]}$ is
generated by the stock-price $S$ and augmented to satisfy the usual
conditions of completeness and right-continuity. $\F=\F_T$, and the 
measure $\P$ is defined by the models in Definition \ref{dfn:models}.

\begin{dfn}\label{dfn:portfolio}
A \emph{portfolio}, or trading strategy, is an
adapted process $\mathbf{\Phi}=(\mathbf{\Phi}_t)_{t\in[0,T]}=
(\Phi^0_t,\Phi_t)_{t\in[0,T]}$, 
where $\Phi^0_t$ denotes the number of bonds and 
$\Phi_t$ denotes the number of shares owned by the investor at time 
$t$.  
The \emph{value} of the portfolio $\mathbf{\Phi}$ at time $t$ is
$$
V_t(\mathbf{\Phi}) = \Phi^0_t + \Phi_t S_t,
$$
since everything is discounted by the bond.
The class of portfolios is denoted by $\A$.
\end{dfn}


There are some slightly different versions of the notion of free lunch,
or arbitrage, that in discrete time would make little or no difference.
However, in continuous time the issue of arbitrage is quite subtle as
can be seen from the fundamental theorem of asset pricing by
Delbaen and Schachermayer \cite[Theorem 1.1]{DelbaenSchachermayer98}.
We use the following definitions:

\begin{dfn}\label{dfn:arbitrage}
\begin{enumerate}
\item
A portfolio $\mathbf{\Phi}$ is \emph{arbitrage} if
$V_0(\mathbf{\Phi})=0$, $V_T(\mathbf{\Phi})\ge 0$ a.s., and
$\P[V_T(\mathbf{\Phi})>0]>0$.
\item
A portfolio $\mathbf{\Phi}$ is \emph{strong arbitrage} if 
$V_0(\mathbf{\Phi})=0$, and there  exists a constant $c>0$ such that
$V_T(\mathbf{\Phi})\ge c$ a.s..
\item 
A sequence of portfolios $(\mathbf{\Phi}^n)_{n\in\N}$ is 
\emph{approximate arbitrage} if $V_0(\mathbf{\Phi}^n)=0$ for all $n$
and 
$
V^\infty_T =\lim_{n\to\infty} V_T(\mathbf{\Phi}^n) 
$
exists in probability, $V^\infty_T \ge 0$ a.s., and $\P[V_T^\infty>0]>0$.
\item
A sequence of portfolios is \emph{strong approximate arbitrage} if it
is approximate arbitrage and there exists a constant $c>0$ such that
$V^\infty_T\ge c$ a.s..
\item
A sequence of portfolios $(\mathbf{\Phi}^n)_{n\in\N}$ is 
\emph{free lunch with vanishing risk} if it is approximate arbitrage, 
and
$$
\lim_{n\to\infty} \esssup_{\omega\in\Omega}
\left|V_T(\mathbf{\Phi}^n)(\omega)
\1_{\left\{V_T(\mathbf{\Phi}^n)<0\right\}}\right| = 0.
$$
\end{enumerate}
\end{dfn}

\section{Trading with (almost) simple strategies}

In this section we consider non-continuous trading in continuous
time.  The basic classes of portfolios are:

\begin{dfn}\label{dfn:si-as}
\begin{enumerate}
\item
A portfolio is \emph{simple} if there exists a finite number of 
stopping times $0\le \tau_0\le\cdots\le\tau_n\le T$ such that the  
portfolio is constant on $(\tau_k,\tau_{k+1}]$, i.e.
$$
\Phi_t = \sum_{k=0}^{n-1} \phi_{\tau_k}\1_{(\tau_k,\tau_{k+1}]}(t),
$$
where $\phi_{\tau_k}\in\F_{\tau_k}$, and an analogous expression holds for $\Phi^0$. The class of simple  portfolios is 
denoted by $\A^{\si}$. 
\item 
A portfolio is \emph{almost simple} if there exists a sequence 
$(\tau_k)_{k\in\N}$ of non-decreasing $[0,T]$-valued stopping times such that 
$\P[\exists_{k \in\N} \;\tau_k=T ]=1$ and the portfolio is constant on 
$(\tau_k,\tau_{k+1}]$, i.e.
$$
\Phi_t = \sum_{k=0}^{N-1} \phi_{\tau_k}\1_{(\tau_k,\tau_{k+1}]}(t),
$$
where $\phi_{\tau_k}\in\F_{\tau_k}$, and $N$ is an a.s. $\N$-valued
random variable, and an analogous expression holds for $\Phi^0$. The class of almost simple portfolios is denoted by 
$\A^{\as}$.
\end{enumerate}
\end{dfn}

\begin{rem}\label{rem:si-as}
Obviously $\A^{\si}\subset\A^{\as}$, and the inclusion is proper.
Also, note that for every $\omega$ the position $\Phi$ is changed
only finitely many times. The difference between $\A^{\si}$ and
$\A^{\as}$ is that in $\A^{\si}$ the number of readjustments
is bounded in $\Omega$, while in $\A^{\as}$
the number of readjustments is generally unbounded.  
\end{rem}

The notion of self-financing is obvious with (almost) simple
strategies:

\begin{dfn}\label{dfn:self-financing-1}
A portfolio $\mathbf{\Phi}\in\A^{\as}$ is \emph{self-financing}
if, for all $k$, its value satisfies
$$
V_{\tau_{k+1}}(\mathbf{\Phi})-V_{\tau_k}(\mathbf{\Phi})
= {\Phi_{\tau_{k+1}}}\left(S_{\tau_{k+1}}-S_{\tau_k}\right),
$$
or, equivalently, the \emph{budget constraint}
$$
\Phi^0_{\tau_{k+1}} + \Phi_{\tau_{k+1}} S_{\tau_{k}} =
\Phi^0_{\tau_{k}} + \Phi_{\tau_{k}} S_{\tau_{k}}
$$
holds for every readjustment time $\tau_k$ of the portfolio.
\end{dfn}

Henceforth, we shall always assume that the portfolios are
self-financing.

\begin{thm}\label{thm:BS-NA-si}
In the BS model there is
\begin{enumerate}
\item
no arbitrage in the class $\A^{\si}$,
\item
strong approximate arbitrage in the class $\A^{\si}$,
\item
strong arbitrage in the class $\A^{\as}$,
\end{enumerate}
\end{thm}

\begin{proof}
The claim (i) follows from the fact that the geometric Bm
remains a martingale in the sub-filtration $(\F_{\tau_k})_{k\le n}$,
and thus the claim reduces to discrete-time considerations. 
Claims (ii) and (iii) follow from the doubling strategy of Example 
\ref{exa:BS-arbitrage-1} below.
\end{proof}

\begin{exa}\label{exa:BS-arbitrage-1} 
Consider, without loss of generality, the risk-neutral normalized
BS model
$$
S_t = s_0 e^{W_t-\frac12t}.
$$
Let $t_k = T(1-2^{-k})$,  
$c_k=e^{\sqrt{T2^{-k}}-\frac12 T2^{-k}}-1$, and
$$
\tau = \inf\left\{ t_k ; 
\frac{S_{t_k}-S_{t_{k-1}}}{S_{t_{k-1}}}\ge c_k \right\}
= \inf\left\{ t_k ; 
\frac{W_{t_k}-W_{t_{k-1}}}{\sqrt{t_{k}-t_{k-1}}} \ge 1
\right\}.
$$ 
Define a self-financing almost simple strategy by setting 
$V_0(\mathbf{\Phi})=0$, and
$$
\Phi_t = \sum_{k=0}^\infty \phi_{t_{k}}
\1_{(t_{k}\wedge \tau,t_{k+1}\wedge\tau]}(t),
$$
where, for $k=0,1,\ldots,$ 
$$
\phi_{t_k}=\frac{1-V_{t_k}(\mathbf{\Phi})}{S_{t_k}c_{k+1}}.
$$
Now, the $c_k$'s were chosen in such a way that $\P[\tau<T]=1$.
So, $\tau=t_N$ a.s. for some random $N\in\N$, and 
\begin{eqnarray*}
V_{\tau}(\mathbf{\Phi}) &=&
V_{t_{N-1}}(\mathbf{\Phi})+
\phi_{t_{N-1}}\left(S_{t_N}-S_{t_{N-1}}\right) \\
&\ge& V_{t_{N-1}}(\mathbf{\Phi})
+\frac{1-V_{t_{N-1}}(\mathbf{\Phi})}{S_{t_{N-1}}c_{N}}
S_{t_{N-1}}c_{N} \\
&=& 1
\end{eqnarray*}
a.s.. So, we have strong arbitrage in the class $\A^{\as}$. Also, 
by setting 
$$
\Phi_t^n = \sum_{k=0}^n \phi_{t_{k}}
\1_{(t_{k}\wedge \tau,t_{k+1}\wedge\tau]}(t),
$$
we see that we have strong approximate arbitrage in the
class $\A^{\si}$.
\end{exa}

In order to exclude doubling-type arbitrage strategies like Example 
\ref{exa:BS-arbitrage-1} one traditionally assumes that the value of the
portfolio is bounded from below:

\begin{dfn}\label{dfn:nds}
A portfolio is \emph{nds-admissible} (\emph{n}o \emph{d}oubling 
\emph{s}trategies) if there exists a constant $a\ge 0$ such that
$$
\inf_{t\in [0,T]} V_t(\Phi) \ge -a \quad \mathrm{a.s.}
$$
The class of nds-admissible portfolios is denoted by $\A^{\nds}$.
\end{dfn}

\begin{rem}\label{rem:nds}
The  sell-short--and--hold strategy 
$\Phi= -\1_{[0,T]}\in \A^{\si}\setminus \A^{\nds}$.
\end{rem}

By Delbaen and Schachermayer \cite[Theorem 1.1]{DelbaenSchachermayer98}
the BS model has no free lunch with vanishing risk, and hence no  
arbitrage, in the class $\A^{\nds}$. The situation for fBS model is
different:

\begin{thm}\label{thm:fBS-arbitrage}
For $H\ne\frac12$, in the fBS model there is
\begin{enumerate}
\item 
free lunch with vanishing risk in the class $\A^{\si}\cap\A^{\nds}$,
\item
strong arbitrage in the class $\A^{\as}\cap\A^{\nds}$.
\end{enumerate}
\end{thm}

\begin{proof}
The claims follow from Cheridito 
\cite[Theorems 3.1 and 3.2]{Cheridito03}.
\end{proof}

Cheridito \cite{Cheridito03} constructed his arbitrage opportunities by 
using the trivial QV of the fBS model ($0$ for $H>1/2$ and $+\infty$ 
for $H<1/2$). So, his constructions do not work in the mfBS model.  
Also, Cheridito's arbitrage strategies are rather implicit in the 
sense that the stopping times they use are not constructed explicitly.  

Let us also note that probably the first one to construct arbitrage in 
the fractional (Bachelier) model was Rogers \cite{Rogers97}. His 
arbitrage was a doubling-type strategy similar to that of Example 
\ref{exa:BS-arbitrage-1} with the twist that he avoided investing on
``bad intervals'' $(t_k,t_{k+1}]$  where the stock price was likely
to fall.  This was possible due to the memory of the fractional
Brownian motion when $H\ne 1/2$.  
With this avoidance he was able to
keep the value of his doubling strategy from falling below any 
predefined negative level, thus constructing an arbitrage opportunity in
the class $\A^{\as}\cap\A^{\nds}$. Let us note that Rogers \cite{Rogers97}
used a representation of the fBm starting from $-\infty$.
So, he used memory from time $-\infty$, while Cheridito \cite{Cheridito03}
and we use memory only from time $0$.

The following very explicit Example \ref{exa:fBS-arbitrage-3}, 
a variant of \cite[Example 7]{BenderSottinenValkeila08}, constructs 
approximate  arbitrage in the fBS model for $H\ne 1/2$ and in the
mfBS model for $H\in(1/2,3/4)$, where the approximating strategies are 
from the class $\A^{\si}$. The construction follows an easy intuition: 
Due to the memory of the fBm the stock price tends to increase (decrease) 
in the future, if it already increased (decreased) in the past if 
$H>1/2$, and the other way around if $H<1/2$. Example 
\ref{exa:fBS-arbitrage-3} also shows that forward integrals with respect to 
fBm with  $H\ne 1/2$ and mixed fBm with $H\in(1/2,3/4)$ are not  
continuous in terms of the integrands. Thus, due to the 
Dellacherie--Meyer--Mokobodzky--Bichteler theorem, this proves that
the fBm is not a semimartingale, and the mixed fBm is not a semimartingale 
when $H\in(1/2,3/4)$.

\begin{exa}\label{exa:fBS-arbitrage-3}
\begin{enumerate}
\item
Consider the fBS model
$$
S_t = e^{ B_t},
$$
where $H\ne1/2$.
Let $t_k^n = T k/n$, $\alpha_H=1$, if  $H>1/2$,
$\alpha_H=-1$, if $H<1/2$, and
$$
\Phi_t^n = \alpha_H n^{2H-1} \sum_{k=1}^{n-1} 
\frac{\log S_{t_k^n}-\log S_{t_{k-1}^n}}{S_{t_k^n}}
\1_{\big(t_k^n,t_{k+1}^n\big]}(t).
$$
Then, assuming $V_0(\mathbf{\Phi}^n)=0$, and applying Taylor's theorem,
\begin{eqnarray*}
V_T(\mathbf{\Phi}^n) &=& \alpha_H n^{2H-1}\sum_{k=1}^{n-1}\left(B_{t_k^n}-B_{t_{k-1}^n}\right)
\left(\frac{S_{t_{k+1}^n}}{S_{t_{k}^n}}-1\right) \\
&=& \alpha_H n^{2H-1}\sum_{k=1}^{n-1}\left(B_{t_k^n}-B_{t_{k-1}^n}\right) \left(B_{t_{k+1}^n}-B_{t_{k}^n}\right)\\ && +  \alpha_H n^{2H-1}\sum_{k=1}^{n-1}  \left(B_{t_k^n}-B_{t_{k-1}^n}\right) e^{\xi^n_k}
\left(B_{t_{k+1}^n}-B_{t_{k}^n}\right)^2,
\end{eqnarray*}
where $|\xi^n_k|\in [0, |B_{t_{k+1}^n}-B_{t_{k}^n}|]$. Now the first term tends to $
T^{2H}\left|2^{2H-1}-1\right|$
in probability by \cite[Theorem 9.5.2]{KarlinTaylor75}, and the second one vanishes as $n$ goes to infinity
using the H{\"o}lder continuity of fBm $B$.

\item Consider the mfBS model
$$
S_t = e^{W_t -\frac12 t + B_t},
$$
where $H\in(1/2,3/4)$.
The strategy of part (i) will still be strong approximate arbitrage.
Indeed, after a Taylor expansion as above, we basically have to deal with the sum of the four terms
\begin{equation}\label{eq:decomp}
\int_0^T K^n_t\d W_t,\; 
\int_0^T L^n_t \d W_t, \;
\int_0^T K^n_t \d B_t, \; 
\int_0^T L^n_t \d B_t, 
\end{equation}
where
\begin{eqnarray*}
K^n_t &=& n^{2H-1}
\sum_{k=1}^{n-1} \1_{\big(t_k^n,t_{k+1}^n\big]}(t)
\left(W_{t_k}-W_{t_{k-1}}\right), \\
L^n_t &=& n^{2H-1}
\sum_{k=1}^{n-1} \1_{\big(t_k^n,t_{k+1}^n\big]}(t)
\left(B_{t_k}-B_{t_{k-1}}\right),
\end{eqnarray*}
and the integrals are just shorthand notation for the forward Riemann sums. Note that $K^n$ and $L^n$ converge to zero uniformly in probability by the H\"older continuity of (fractional) Brownian motion for $H<3/4$. Therefore, the first two terms in (\ref{eq:decomp})
will tend to zero in probability by the 
Dellacherie--Meyer--Mokobodzky--Bichteler theorem \cite[Theorem II.11]{Protter04}.
The third term will tend to zero in probability because of the
independence of $W$ and $B$. The fourth term will go to 
$T^{2H}(2^{2H-1}-1)$ in probability by part (i) of this example. We also note that $\Phi^n S$ inherits the unform convergence in probability to zero from $K^n+L^n$. Hence the amount of money invested in the stock converges to zero as $n$ tend to infinity. 
\end{enumerate}
\end{exa}

For the mfBS model the situation is the following:

\begin{thm}\label{thm:mfBS-as-si}
For the mfBS model there is
\begin{enumerate}
\item
strong approximate arbitrage in the class $\A^{\si}$
if $H\in(1/2,3/4)$,
\item
no free lunch with vanishing risk in the class $\A^{\nds}$
if $H\in (3/4,1)$.
\end{enumerate}
\end{thm}

\begin{proof}
Claim (i) follows from Example \ref{exa:fBS-arbitrage-3}(ii). Claim (ii) 
follows from Cheridito \cite{Cheridito01}.  Indeed, in 
\cite{Cheridito01} it is shown that in this case the mixed fBm is 
actually equivalent in law to a Bm.
\end{proof}

Although the situation is bad arbitrage-wise for the fBS and the mfBS 
models in the class $\A^{\si}\cap\A^{\nds}$, Cheridito \cite{Cheridito03}
showed that there is no arbitrage in the fBS model if there must be a 
fixed positive time between the readjustments of the portfolio (later 
arbitrage in this class was studied by Jarrow et al. 
\cite{JarrowProtterSayit09}): 

\begin{dfn}\label{dfn:T-si}
Let $\T$ be a class of finite sequences of non-decreasing stopping times 
$\bm{\tau}=(0\le \tau_0\le \cdots \le \tau_n\le T)$
satisfying some additional conditions, which can be specified as in Proposition \ref{pro:fBS-h-si} or Definition \ref{dfn:delay-simple} below. A simple portfolio $\Phi$ is 
\emph{$\T$-simple} if it is of the form
$$
\Phi_t = \sum_{k=0}^{n-1} \phi_{\tau_k}\1_{(\tau_k,\tau_{k+1}]}(t),
$$
where $\phi_{\tau_k}\in \F_{\tau_k}$, $\bm{\tau}=(\tau_k)_{k=0}^n\in \T$.
The class of $\T$-simple strategies is denoted by $\A^{\T-si}$.
\end{dfn}

\begin{pro}\label{pro:fBS-h-si}
Let 
$
\T_h = \cup_{h>0}\left\{ \bm{\tau} ;
\tau_{k+1}-\tau_k \ge h\right\}.
$
Then
there is no arbitrage in the fBS model in the class $\A^{\T_h-\si}$.
\end{pro}

\begin{proof}
The claim is Cheridito's \cite[Theorem 4.3]{Cheridito03}.
\end{proof}

\section{Trading with delay-simple strategies}

While Proposition \ref{pro:fBS-h-si} seems promising the class
$\A^{\T_h-\si}$ is more restrictive than it may appear at a first
sight. Indeed, e.g. the archetypical stopping time 
$\tau=\inf\{ t\ge 0; S_t -S_0 \ge 1\}$ does not belong to $\T_h$ if
$S$ is the geometric Bm. To remedy this problem we propose the 
following more general class of stopping times and simple strategies:

\begin{dfn}\label{dfn:delay-simple}
\begin{enumerate}
\item
For any stopping time $\tau$,
let $C_{S_{\tau}}^+([\tau,T])$ be the random space of continuous positive
paths $\omega=(\omega_t)_{t\in[\tau(\omega),T]}$ with 
$\omega_{\tau(\omega)}=S_{\tau(\omega)}(\omega)$ fixed. 
A sequence of non-decreasing stopping times $\bm{\tau}=(\tau_k)_{k=0}^n$
satisfies the \emph{delay} property if for all $\tau_k$ there is an
$\F_{\tau_k}$-measurable open \emph{delay set} 
$U_k\subset C_{S_{\tau_k}}^+([\tau_k,T])$ and an $\F_{\tau_k}$-measurable 
a.s. positive random variable $\eps_k$ such that 
$\tau_{k+1}-\tau_k\ge \eps_k$ in the set 
$U_k\cap\{\tau_{k+1}>\tau_k\}$.
The set of non-decreasing sequences of stopping times satisfying the 
delay property is denoted by $\T_{\de}$.
\item
The class of \emph{delay-simple} strategies is $\A^{\T_{\de}-\si}$.
\end{enumerate}
\end{dfn}

\begin{thm}\label{thm:na-delay}
All the models BS, fBS and mfBS are free of arbitrage in the
class $\A^{\T_{\de}-\si}$. 
\end{thm}

Before we prove Theorem \ref{thm:na-delay} we discuss the class
of delay-simple strategies.

\begin{rem}\label{rem:h-vs-de}
The difference between the classes $\T_h$ and $\T_{\de}$ is that
in $\T_h$ there is a fixed delay $h>0$ between the stopping times,
while in $\T_{\de}$ the delay between the stopping times depend
on the path one is observing:  If there is a delay on the path
you are observing then there is also a delay on all the paths that
are close enough of the path that one is observing.

Obviously $\T_h\subset \T_{\de}$, and the inclusion is proper.
\end{rem}

\begin{exa}\label{exa:de-si}
The following sequences of stopping times are in $\T_{\de}$: 
\begin{enumerate}
\item
$$
\tau_{k+1} = \inf\left\{ t> \tau_{k} ; S_t -S_{\tau_{k}}\ge b_t^k\right\},
$$
where $b^k$'s are continuous function with $b_{\tau_k}^k>0$.
Indeed, take
$$
U_k = \left\{ \omega; S_t(\omega) < \omega^0_t \mbox{ for all } 
t\in [\tau_{k},T]\right\},
$$
where $\omega^0$ is some path for which 
$\tau_{k+1}(\omega^0)>\tau_{k}(\omega^0)$.
\item
$$
\tau_{k+1} = \inf\left\{ t> \tau_{k} ; S_t -S_{\tau_{k}}\le a_t^k\right\},
$$
where $a^k$'s are continuous function with $a_{\tau_k}^k<0$.
Indeed, take
$$
U_k = \left\{ \omega; S_t(\omega) > \omega^0_t \mbox{ for all } 
t\in [\tau_{k},T]\right\},
$$
where $\omega^0$ is some path for which 
$\tau_{k+1}(\omega^0)>\tau_{k}(\omega^0)$.
\item
One can show that 
$$
\tau_{k+1} = \inf\left\{ t > \tau_{k};
S_t-S_{\tau_{k}} \le a_t^k \mbox{ or } S_t-S_{\tau_{k}}\ge b_t^k\right\},
$$
$a^k$'s and $b^k$'s are continuous with $a_{\tau_k}^k <0<b_{\tau_k}^k$,
is in $\T_{\de}$ (see \cite[Example 6 (i)]{BenderSottinenValkeila08}).
\end{enumerate} 
\end{exa}

\begin{exa}\label{exa:no-de-si}
We construct a stopping time $\tau$ in the fractional Wiener space such that $(\tau_0,\tau_1):=(0,\tau)$ is
not in $\T_{\de}$: $\tau=\inf\{ t>0; e^{B_t+t^a}=1\}$. By the law of
iterated logarithm $\tau>0$ a.s. if $a<H$. 
However, any open set $U\subset C_{S_0}^+([0,T])$ contains sequences
$(\omega^n)$ for which $\tau(\omega^n)\to 0$.
\end{exa}

\begin{dfn}\label{dfn:T-upndown}
A process $S$ satisfies the \emph{$\T$-conditional up'n'down}
property ($\T$-CUD) if, for all $\bm{\tau}\in \T$ and all $k$, either
$$
\P\left[S_{\tau_{k+1}}>S_{\tau_{k}}\big| \F_{\tau_k}\right] >0
\mbox{ and }
\P\left[S_{\tau_{k+1}}<S_{\tau_{k}}\big| \F_{\tau_k}\right] >0
$$
or
$$
\P\left[S_{\tau_{k+1}}=S_{\tau_{k}}\big| \F_{\tau_k}\right] =1.
$$ 
If there are no additional restrictions for $\T$ (except
that it contains non-decreasing finite sequences of stopping times), we say simply
that $S$ satisfies CUD.
\end{dfn}

The following lemma can be proved analogously to  \cite[Lemma 1]{JarrowProtterSayit09}. 
\begin{lmm}\label{lmm:T-upndown}
There is no arbitrage in the class $\A^{\T-\si}$ if and only if
the model satisfies $\T$-CUD.
\end{lmm}

CUD is related to the support of the stock-price model $S$.  Another 
support-related condition we need is:

\begin{dfn}\label{dfn:CFS}
A continuous positive process $S$ has \emph{conditional full support} 
(CFS) if, for all stopping times $\tau$, 
$$
\mbox{supp }\P[S\in\,\cdot\, | \F_{\tau}] = C_{S_{\tau}}^+([\tau,T])
\quad\mbox{a.s.}.
$$
\end{dfn}

\begin{rem}\label{rem:CFS}
\begin{enumerate}
\item
CFS is equivalent to the \emph{conditional small-ball property}: For every stopping time $\tau$, all the
open balls contained in $C_{S_{\tau}}^+([\tau,T])$ have a.s. positive 
regular conditional probability, i.e.
$$
\P\left[ \sup_{t\in[\tau,T]} \left|S_t-S^0_t\right|\le \eps \bigg|
\F_{\tau}\right] > 0
$$
a.s. for all $S^0\in C_{S_{\tau}}^+([\tau,T])$ and 
$\F_{\tau}$-measurable a.s. positive random variables $\eps$.
For a proof of this see Pakkanen \cite[Lemma 2.3]{Pakkanen09}.
\item
By Pakkanen \cite[Lemma  2.10]{Pakkanen09} a process $X$ has CFS with respect 
to its own filtration $\mathcal{F}^X_t=\sigma(X_s,\;0\leq s\leq t)$ if and only if it has the CFS with respect
to the augmentation of $\mathcal{F}^X_t$.
\item
By Guasoni et al. \cite[Lemma 2.9]{GuasoniRasonyiSchachermayer08} one can
replace the stopping times with deterministic times in Definition 
\ref{dfn:CFS}. 
\item 
CFS is neither  necessary nor sufficient for no-arbitrage in $\A^{\si}$. On the one hand, any bonded martingale satisfies no-arbitrage in $\A^{\si}$, but violates CFS. On the other hand
$W_t+t^a$, $a<1/2$, has arbitrage in $\A^{\si}$ by the law of the iterated logarithm, but satisfies CFS.
However, CFS is sufficient for absence of arbitrage with the class $\A^{\T_{\de}-\si}$. This will be shown in the next lemma.
\end{enumerate}
\end{rem}

\begin{lmm}\label{lmm:CFS-delay}
Suppose $S$ has CFS. Then there is no arbitrage in the model $S$
in the class $\A^{\T_{\de}-\si}$.
\end{lmm}

\begin{proof}
By Lemma \ref{lmm:T-upndown} we need to show that the $\T_{\de}$-CUD is 
satisfied. If  $\tau_{k+1}=\tau_k$, this is certainly the case.  So, we 
can assume that $\tau_{k+1}>\tau_k$. 

We show that $\P[S_{\tau_{k+1}}>S_{\tau_k}|\F_{\tau_k}]>0$ a.s.; the 
proof for $\P[S_{\tau_{k+1}}<S_{\tau_k}|\F_{\tau_k}]>0$ a.s. follows
analogously.

By the CFS it is enough to show that  
$\{ S_{\tau_{k+1}}>S_{\tau_k}\}\subset C_{S_{\tau_k}}^+([\tau_k,T])$
contains an open set. Let $U_k$ be an $\eps_k$-delay set for $\tau_k$, i.e. $U\subset C_{S_{\tau_k}}^+([\tau_k,T])$ is open and $\tau_{k+1}-\tau_k\geq \epsilon_k$ on $U_k$.
We first assume that $U_k$ contains a strictly increasing path $\omega^0$ on $[\tau_k,T]$. Denote by $B_{\omega^0}(\eta_k)$ the open $\eta_k$-ball around $\omega_0$ in $C_{S_{\tau_k}}^+([\tau_k,T])$. Choosing $\eta_k$ sufficiently small we have $B_{\omega^0}(\eta_k)\subset U_k$ (because $U_k$ is open)
and $\omega^0_{\tau_k+\epsilon_k}> \omega^0_{\tau_k}+\eta_k$ (because $\omega^0$ is strictly increasing). Hence, for every $\omega\in B_{\omega^0}(\eta_k)$,
\begin{eqnarray*}
\omega_{\tau_{k+1}(\omega)}-S_{\tau_{k}}
&>& \omega^0_{\tau_{k+1}(\omega)} -\eta_k -S_{\tau_k} \\
&\ge& \omega^0_{\tau_{k}+\eps_k} -S_{\tau_k} -\eta_k \\
&=& \omega^0_{\tau_k+\eps_k}-\omega^0_{\tau_k}-\eta_k \\
&>& 0, 
\end{eqnarray*}
So, $B_{\omega^0}(\eta_k)\subset\{ S_{\tau_{k+1}}>S_{\tau_k}\}$, and the claim follows, if $U_k$ contains a strictly increasing path.
If $U_k$ does not contain a strictly increasing path, we proceed as follows. Being an open set in $C_{S_{\tau_k}}^+([\tau_k,T])$, $U_k$ contains paths
that are strictly increasing on a small enough interval $[\tau_k,\tau_k+2\eta_k]$. Hence, there is a strictly increasing path $\omega^0$ and an open  ball $B_k$ around $\omega^0$ in $C_{S_{\tau_k}}^+([\tau_k,T])$ such that any $\omega\in B_k$ coincides with some path $\bar\omega\in U_k$ on the segment $[\tau_k,\tau_k+\eta_k]$. Hence, $\tau_{k+1}(\omega)-\tau_k\geq (\tau_{k+1}(\bar \omega)-\tau_k)\wedge \eta_k \geq \epsilon_k\wedge \eta_k=:\epsilon^0_k$ for every $\omega \in B_k$. Therefore $B_k$ is an $\epsilon^0_k$-delay set which contains a strictly increasing path and so the first case applies. 
\end{proof}

\begin{proof}[Proof of Theorem \ref{thm:na-delay}]
By \cite[Theorem 2.1]{GasbarraSottinenVanZanten08} the Bm, the fBm, and
the mixed fBm all have CFS in the space $C_0([0,T])$ (with respect to the filtration generated by the respective process), since their
spectral measures have heavy enough tails. 
For a nice proof that fBm has CFS see also \cite{Cherny08}. 
So, the BS, the fBS, and the 
mfBS models all have CFS in $C_{s_0}^+([0,T])$, because with any 
homeomorphism $\eta$ on $C_0([0,T])$ the mapping 
$\omega\mapsto s_0e^{\omega+\eta}$ is a homeomorphism between  
$C_0([0,T])$ and $C_{s_0}^+([0,T])$. 
So, the claim follows from Lemma \ref{lmm:CFS-delay}.
\end{proof}

\section{Continuous trading}

While the previous sections were concerned with trading strategies which can be readjusted finitely many times only, we will now admit continuous readjustment of the portfolio. A natural generalization of the self-financing property in Definition \ref{dfn:self-financing-1} can be given in terms of forward integrals. Here we stick to the simplest possible definition of forward integrals due to \cite{Foellmer81}, but refer to \cite{RussoVallois93} for the general theory.

\begin{dfn} \label{dfn:forward}
Let $t\leq T$ and
let $X=(X_s)_{s\in[0,T]}$ be a continuous process. The
\emph{forward integral} of a process $Y=(Y_s)_{s\in[0,T]}$ with
respect to $X$ (along dyadic partitions) is
$$
\int_0^t Y_s\, \d X_s := \lim_{n\to\infty}
\sum_{i=0,\ldots, 2^n-1,\atop Ti/2^n \le t} Y_{Ti/2^n} \left(
X_{T(i+1)/2^n}-X_{Ti/2^n}\right),
$$
if the limit exists $\P$-almost surely.
\end{dfn}
If necessary, we interpret the forward integral in an improper sense at $t=T$. It\^o's formula for the forward integral depends on the quadratic variation of the integrator.

\begin{dfn}\label{dfn:QV}
The pathwise \emph{quadratic variation} (QV) of a stochastic process (along dyadic partitions)
is 
$$
\la X \ra_t :=  \lim_{n\to\infty} \sum_{i=0,\ldots, 2^n-1,\atop Ti/2^n \le t} \left(X_{T(i+1)/2^n}-X_{Ti/2^n}\right)^2,
$$
if, for all
$t\le T$, the limit exists $\P$-almost surely.
\end{dfn}

\begin{pro}\label{pro:QV-mfBS}
\begin{enumerate}
\item For the fBS model and the mfBS model with $H<1/2$ the limit in Definition \ref{dfn:QV} diverges to infinity.
\item For the fBS model with $H>1/2$, the QV is constant 0. 
\item
The QV in the BS model and in the mfBS model with $H>1/2$ is given by
$$
\d\la S\ra_t = \sigma^2 S_t^2\d t.
$$
\end{enumerate}
\end{pro}
\begin{proof}
It is well known that Bm has the identity map as QV. Moreover, fBm has zero quadratic variation for $H>1/2$ and infinite quadratic variation for $H<1/2$, see e.g. \cite{BiaginiHuOksendalZhang08}, Chapter 1.8. By independence, the QV of the mixed fBm is the sum of the QV of Bm and fBm. Finally, the stock models under consideration are $\mathcal{C}^1$-functions of these processes (up to a finite variation drift), and so a result by \cite{Foellmer81}, p. 148, applies.
\end{proof}

The following It\^o formula for the forward integrals with continuous integrator can be derived by a Taylor expansion as usual, see \cite{Foellmer81}.

\begin{lmm}\label{lmm:Ito}
Let $X$ be a continuous process with continuous QV.  Suppose
$f\in\mathcal{C}^{1,2}([0,T]\times\R$). Then, for $0 \le
t\leq T$,
\begin{eqnarray*}
f(t,X_t) &=& f(0,X_0)
+ \int_0^t \frac{\p}{\p t}f(u,X_u)\, \d u 
 + \int_0^t \frac{\p}{\p x}f(u,X_u) \, \d X_u \\
& & + \frac{1}{2}\int_0^t \frac{\p^2}{\p
x^2}f(u,X_u)\,
\d\la X \ra_u 
\end{eqnarray*}
In particular, this formula implies that the forward integral on the
right hand side exists and has a continuous modification.
\end{lmm}

After this short digression on forward integrals we can introduce several classes of portfolios.
\begin{dfn}
\begin{enumerate}
 \item A portfolio is \emph{self-financing}, if, for all $0\leq t \leq T$,
$$
V_t({\bf \Phi})=V_0({\bf \Phi})+\int_0^t \Phi_t dS_t.
$$
The class of self-financing portfolios (without any extra constraints) is denoted by $\mathcal{A}$.
\item A self-financing portfolio is called a \emph{spot strategy}, if $\Phi_t=\varphi(t,S_t)$ for some deterministic function $\varphi$, i.e. the number of shares held in the stock depends on time and the spot only. We apply the notation $\mathcal{A}^{spot}$ for the class of spot strategies. 
\end{enumerate}
 \end{dfn}

The following theorem discusses arbitrage with spot strategies in the BS model. It again illustrates some subtleties of arbitrage theory in continuous time, even for models which admit an equivalent martingale measure. As in the case of almost simple strategies, arbitrage is possible, if arbitrarily large losses are allowed prior to maturity.

\begin{thm}\label{thm:BS-arbitrage-2}
\begin{enumerate}
 \item In the BS model  there is strong arbitrage in the class $\mathcal{A}^{spot}$.
\item In the BS model  there is no free lunch with vanishing risk in the class $\mathcal{A}\cap\mathcal{A}^{nds}$.
\end{enumerate}
\end{thm}
\begin{proof}
(i) We give a direct construction making use of It\^o's formula (Lemma \ref{lmm:Ito}) and the QV of the Black-Scholes model. W.l.o.g. we assume $\sigma=1$ and $\mu=0$.
Let
$$
\Phi_t = -\frac{\frac{\p}{\p x}v(t,W_t)}{S_t},
$$
where $v(t,x)$ is the heat kernel
$$
v(t,x) = \frac{1}{\sqrt{2\pi(T-t)}}e^{-\frac{1}{2}\frac{x^2}{T-t}}.
$$
By Lemma \ref{lmm:Ito}, applied to the Bm $W$,
$$
V_T(\Phi) = \int_0^T \Phi_t \d S_t = - \int_0^T \frac{\p}{\p x}v(t,W_t) \d W_t=v(0,0)-v(T,W_T) = \frac{1}{\sqrt{2\pi T}},
$$
almost surely. So, we have constructed a strong arbitrage and it belongs to the class $\A^{spot}$, because the Bm $W$ is a deterministic function of time and the Black-Scholes stock $S$. \\
(ii) The BS model has an equivalent martingale measure. Hence the fundamental theorem of asset pricing \cite{DelbaenSchachermayer94} ensures that there is no free lunch with vanishing risk with nds-admissible, self-financing strategies.
\end{proof}

The construction of the `doubling' arbitrage in the previous theorem only relied on the quadratic variation structure of the model. In the pure fractional BS model with $H>1/2$ the QV is constant zero. This fact, combined with It\^o's formula, can be exploited to construct an nds-admissible arbitrage in class $\mathcal{A}^{spot}$. The following simple example is due to Dasgupta and Kallianpur \cite{DasguptaKallianpur00} and Shiryaev \cite{Shiryaev98}.

\begin{exa}\label{ex:fBS-smootharbitrage}
Choosing $\Phi_t=S_t-S_0$, we obtain by It\^o's formula (Lemma \ref{lmm:Ito}) and the zero QV property of the fBS model with $H>1/2$,
$$
(S_t-S_0)^2=2\int_0^t \Phi_u \d S_u.
$$
Hence, ${\bf \Phi}$ is nds-admissible (it is bounded from below by 0) and an arbitrage. Again, this construction of an arbitrage applies to all models with zero QV and $P(S_T\neq S_0)>0$.
\end{exa}

We now consider hedging in the fBS model with Hurst parameter larger than a half. 
Although there exists strong arbitrage
in the class  $\A^{\nds}\cap\A^{\as}$ by Theorem \ref{thm:fBS-arbitrage}, 
one can still consider the hedging problem in the fBS model. Indeed, in 
spite of arbitrage one may still be interested in hedging per se.  But, 
it must be noted that hedging cannot be used as a pricing paradigm in the
presence of strong arbitrage, since for any hedge one can find a 
super-hedge with smaller initial capital by combining the hedge with a 
strong arbitrage. 

By a straightforward generalization of the previous example, we observe that a smooth European style option, i.e. with pay-off $f(S_t)$ for some $f\in \mathcal{C}^1$ can be hedged with initial endowment $f(S_0)$ and the strategy $\Phi_t=f'(S_t)$. In reality many options, like vanilla options, have a convex pay-off function, which does not belong to class $\mathcal{C}^1$. A generalization to this situation is possible with some extra effort as outlined next.

\begin{dfn}
Let $f: \R _+ \to \R_+$ be a convex function and $H > 1/2$. If we can 
find a self-financing strategy $\mathbf{\Phi}$ and a constant $c^f$ such 
that 
\begin{equation}\label{eq:fbs-hedging}
f(S_T) = c^f + \int _0^T \Phi _s \,\d S_s ,
\end{equation}
then $\mathbf{\Phi}$ is a \emph{hedging strategy} and $c^f$ is a 
\emph{hedging cost} of the option $f(S_T)$.
\end{dfn}

\begin{rem}
\begin{enumerate}
\item Because of the strong arbitrage possibilities in the fBS model one
cannot interpret the hedging cost $c^f$ as a minimal super-replication 
price.
\item  The strong arbitrage possibility of the fBS model does not imply 
that one can take $c^f=0$ in (\ref{eq:fbs-hedging}): One can super-hedge
with zero capital, but the hedge may not be exact.  While from the purely
monetary point of view this does not matter, there may be situations
where one is penalized for not hedging exactly.
\end{enumerate}
\end{rem}

If $f$ is a convex function, then $f^+_x $ $(f^-_x)$ is the right (left) 
derivative of $f$. The following theorem can be regarded as a generalization of the It\^o formula in Lemma \ref{lmm:Ito} for non-smooth convex functions in the pure fractional Brownian motion setting.

\begin{thm}\label{t:fbs-hedging}
Suppose $S$ is the fBS model with $H>1/2$ and $f$ is a convex function. Then 
\begin{equation}\label{eq:fbs-hedge}
 f(S_T) = f(S_0 ) + \int _0 ^T f^+ _x(S_u) \, \d S_u.
\end{equation}
In particular, the European option $f(S_T)$ can be perfectly hedged with cost $f(S_0)$ and the hedging strategy given by $\Phi_t=f^+_x(S_t)$.
\end{thm}
\begin{proof}
One proves Theorem \ref{t:fbs-hedging} by showing that the integral 
exists as a generalized Lebesgue--Stieltjes integral. This is done with 
the help of some fractional Besov space techniques. Finally, one proves 
that the integral exists as a forward integral and actually even as a Riemann-Stieltjes integral. For the rigorous proof
see \cite{AzmoodehMishuraValkeila09}. Note that one can replace the right derivative  $f^+ _x$ by the left derivative $f^- _x$, as both derivatives differ on a countable set only.
\end{proof}

\begin{exa}\label{ex:fbs-arbitrage} If the convex function $f$ corresponds to the call option, i.e. 
$f(x) = (x-K)^+$, then we observe that
the stop--loss--start--gain portfolio replicates the call option:
$$
(S_T -K)^+ = (S_0-K)^+ +\int _0^T \1_{\{ S_t \ge K\}} \,\d S_t.
$$
Note that this again gives an arbitrage strategy, if the option is 
at--the--money or out--of--the--money.
\end{exa}

If $H<1/2$  stochastic integrals for typical spot strategies with respect to the fBS model fail to exist.  So it makes little sense to consider 
continuous trading in this situation. This unfortunate property is related to the infinite QV of the fBS model for small Hurst parameter and thus applies for the mixed model with $H<1/2$ as well. 

For the remainder of the section we shall therefore discuss the mfBS model with $H>1/2$. In the case $H>3/4$, the mfBS model is equivalent in law to the BS model, see \cite{Cheridito01}. Therefore, all constructions of arbitrages with doubling strategies and all results on no arbitrage with nds-strategies directly transfer from the BS model to the mfBS model with $H>3/4$. Moreover, the latter model inherits the completeness of the BS model. We now discuss to what extent the mixed model with $1/2<H\leq 3/4$ differs from the BS model. The argumentation below only makes use of the fact that the mixed model has the same QV as the BS model and has conditional full support.

\begin{thm}\label{thm:mixedBSV}
 Suppose $S$ is the mfBS with $H>1/2$. Then,
\begin{enumerate}
 \item There is strong arbitrage in the class $\mathcal{A}^{spot}$.
 \item There is no nds-admissible arbitrage ${\bf \Phi}$ of the form 
$$
\Phi_t=\varphi\left(t,\max_{0\leq u \leq t} S_u, \min_{0\leq u \leq t} S_u, \int_0^t S_u \d u,S_t\right)
$$
with $\varphi\in \mathcal{C}^{1}([0,T]\times \mathbb{R}_+^4)$. A strategy of this form will be called \emph{smooth} from now on.
\end{enumerate}
\end{thm}
\begin{proof}
 (i) Here the same constructive example as in Theorem \ref{thm:BS-arbitrage-2} applies, because the mfBS model has the same QV as the BS model. \\
(ii) We fix some smooth strategy ${\bf \Phi}$. By a slightly more general It\^o formula than the one in Lemma \ref{lmm:Ito} one can conclude that there is a continuous functional $v:[0,T]\times C_{s_0}^+([0,T]) \rightarrow \mathbb{R}$ such that $V_t({\bf \Phi})=v(t,S)$. By the full support property, the paths of the mfBS model can be approximated by paths of the BS model and vice versa. In this way, absence of arbitrage can be transferred from the BS model to the mfBS model. The details are spelled out in \cite{BenderSottinenValkeila08}, Theorem 4.4.

We point out, that in the special case  ${\bf \Phi}=(\Phi^0,\Phi)\in\A$ with $\Phi_t=\varphi(t,S_t)$ and $\Phi^0_t=\varphi^0(t,S_t)$ for some sufficiently smooth functions $(\varphi,\varphi^0)$, the value process $V_t({\bf \Phi})$ can be linked to a PDE. This was exploited in \cite{AndroshcukMishura06} in order to prove absence of arbitrage in this special case.
\end{proof}

\begin{rem}
\begin{enumerate}
 \item In Theorem \ref{thm:mixedBSV}, (ii), the differentiability of $\varphi$ at $t=T$ can be relaxed to some extent and absence of arbitrage still holds. The resulting class of strategies contains hedges for many relevant European, Asian, and lookback options. These hedges (as functionals on the paths) and the corresponding option prices (deduced by hedging and no-arbitrage relative to this class of portfolios) are the same as in the BS model. For the details we refer to \cite{BenderSottinenValkeila08}. We note that this robustness of hedging strategies was already shown by Schoenmakers and Kloeden \cite{SchoenmakersKloeden99} in the case of European options.
\item The no-arbitrage result in Theorem \ref{thm:mixedBSV}, (ii), can be extended in several directions. Additionally to the running maximum, minimum and average, the strategy can depend on other factors, which are supposed to be of finite variation and satisfy some continuity condition as functionals on the paths. The investor also is allowed to switch between different smooth strategies at a large class of stopping times and still absence of arbitrage holds true for these stopping-smooth strategies. For the exact conditions on the stopping times we refer to Section 6 in \cite{BenderSottinenValkeila08}, but we note that many typical ones such as the first level crossing of the stock are included.
\end{enumerate}
\end{rem}

\section{Trading under transaction costs}

Recently Guasoni \cite{Guasoni06} and Guasoni et. al. \cite{GuasoniRasonyiSchachermayer08} have shown, that allowing transaction costs in the fBS model the arbitrage possibilities disappear. First they introduce, following Jouini and Kallal \cite{JouiniKallal95},  the notion of $\epsilon$- consistent price system.

\begin{dfn}
 Let $S$ be a continuous process with paths in $C^+ _{S_0}([0,T]) $. An $\epsilon $-consistent price system is 
a pair $(\widetilde S, \Q)$ of a probability $\Q$ equivalent to $\P$, and a $\Q$-martingale $\widetilde S =(\widetilde S_t ) _{0\le t\le T} $, such that $S_0= \widetilde S_0 $, and for $0\le t\le T$, $\epsilon > 0$ 
$$
1-\epsilon \le \frac{\widetilde S _t}{S_t } \le 1+ \epsilon,  \mbox{ a.s.} 
$$
\end{dfn}

With proportional transaction costs one can not use continuous trading. Denote by $\mathcal V(\Phi )$ the total variation of the process $\Phi$.  In this section a trading strategy $\Phi $ is predictable finite-variation $\R$-valued process such that $\Phi _0 = \Phi_T=0$. The value of $\Phi$ with $\epsilon$- costs  $V^\epsilon (\Phi )$ is 
$$
V^\epsilon  (\Phi) = \int _0^T \Phi _s dS_s - \epsilon \int_0^T S_s d \mathcal V(\Phi) _ s .
$$
Define $V^\epsilon _t (\Phi ) $ by 
$$
V^\epsilon _t (\Phi ) = V^\epsilon (\Phi 1_{(0,t)}) ,
$$
and so $V^\epsilon (\Phi) = V ^\epsilon _T (\Phi )$. 

Next, we define the set of admissible strategies in this context, following \cite{GuasoniRasonyiSchachermayer08}:
given $M> 0$, the strategy $\Phi$ is $M$-admissible, if for all $t\in [0,T]$ we have that 
$$
V^\epsilon _t(\Phi) \ge -M (1+S_t) \mbox{ a.s.} 
$$
The set of $M$-admissible strategies is denoted by $\mathcal A ^{\mbox{adm}}_M (\epsilon) $. Define also 
$$
\mathcal A ^{\mbox{adm}}(\epsilon) = \cup _{M>0} \mathcal A   ^{\mbox{adm}}_M (\epsilon) .
$$
Finally we say that $S$ admits arbitrage with $\epsilon$-transaction costs if there is $\Phi \in \mathcal A ^{\mbox{adm}}(\epsilon ) $  such that $V^\epsilon (\Phi) \ge 0$ and $\P ( V^\epsilon (\Phi) > 0) > 0 $.  

We can now state the fundamental theorem of asset pricing with $\epsilon$-transaction costs given in \cite[Theorem 1.11]{GuasoniRasonyiSchachermayer08}: 

\begin{thm}
 Let $S\in C_{s_0}^+([0,T]) $. Then the following two conditions are equivalent:
\begin{enumerate}
 \item For each $\epsilon > 0 $ there exists an $\epsilon$-consistent price system.
\item For each $\epsilon > 0 $, there is no arbitrage for $\epsilon$-transaction costs. 
\end{enumerate}
\end{thm}

\cb{It is shown by Guasoni et al. \cite{GuasoniRasonyiSchachermayer08a} that conditional full support implies the existence of an  $\epsilon$-consistent price system for every $\epsilon>0$. Therefore, the fBS models and the mfBS models do not adimit arbitrage under transaction cost with the classes of strategies $\mathcal A ^{\mbox{adm}}(\epsilon)$ for $\epsilon>0$.}

We will study a concrete hedging problem with proportional transaction costs. 

In Theorem \ref{t:fbs-hedging} it was shown that the European option $f(S_T)$ can be perfectly hedged with cost $f(S_0)$ and hedging strategy $\Phi _t = f^- _x(S_t )$.  Take $T=1$, put $t^n _ i = \frac{i}{n}$, $i=0, \dots , n$,  and consider the discretized hedging strategy $\Phi ^n $
\begin{equation}\label{eq:discrete-fBS-hedge}
 \Phi ^n _t = \sum _{i=1}^n f^-_x(S_{t^n_{i-1}} ) 1 _{(t^n _{i-1}, t^n_i ] } (t) .
\end{equation}
 
Consider now discrete hedging with proportional transaction costs $k_n= k_0 n^{-\alpha}$ with $\alpha > 0, k_0> 0$. 
The value of the strategy $\Phi^n $ at time $T=1 $ is 
\begin{equation}\label{eq:transaction-fBS}
V_1 (\Phi ^n; k_n ) = f(S_0) + \int _0^1 \Phi ^n _t dS_t - k_n \sum _{i=1}^n S_{t^n_{i-1}}| f^-_x(S_{t^n_i} ) -
f^-_x(S_{t^n_{i-1}}) | .
\end{equation}
Note that there is no transaction costs at time $t=0$. 

In the next theorem $\mu^f$ is the second derivative $f_{xx}$ of the convex function $f$. The derivative exists in a distributional sense, and $\mu ^f$ is a Radon measure. The occupation measure $\Gamma _{B^H }$ of fractional Brownian motion $B^H$ is defined by 
$\Gamma _{B^H} ([0,t]\times A ) = \lambda \{ s\in [0,t]: B^H_s \in A \} $; here   $\lambda $ is the Lebesgue measure and $A$ is a Borel set.  Denote by $l^H (x, t )$ the local time of fractional Brownian motion $B^H$; recall that local time $l^H$ is the density of the occupation measure with respect the Lebesgue measure. 

The following theorem is proved in \cite{Azmoodeh09}:

\begin{thm}\label{thm:fBS-transactions}
Let $V_1(\Phi  ; k_n) $ be the value of the discrete hedging strategy $\Phi^n$ with proportional
transaction costs $k_n = k_0 n^{-\alpha}$. 
\begin{enumerate}
 \item If $\alpha > 1- H $, then as $n\to \infty $
$$
V_1(\Phi ^n ; k_n ) \to f(S_1) \mbox{ in probability.}
$$
\item If $\alpha = 1-H $, then as $n\to \infty $
\begin{equation}\label{eq:trans-limit}
V_1(\Phi ^n ; k_n ) \to f(S_1) - \sqrt{\frac{2}{\pi}} k_0 \int _\R \int _0^1 S_t dl^H(\ln (a), t) \mu ^f (da) .
\end{equation}
\end{enumerate}

\end{thm}

\begin{rem}
Note that one can write the limit result in \eqref{eq:trans-limit} as 
$$
f(S_1) = f(S_0) + \int _0 ^1 f^-_x(S_u)dS_u + \sqrt{\frac{2}{\pi}} k_0 \int _\R \int _0^1 S_t dl^H(\ln (a), t) \mu ^f (da);
$$ 
if $l^W $ is the local time for Brownian motion, then the It{\^o}-Tanaka formula gives
$$
f(W_1) = f(0) + \int _0 ^1 f^-_x(W_u)dW_u + \frac12 \int _0^1 \int _\R  dl^W(a, u )\mu ^f(da) .
$$
Hence asymptotical transaction costs with $\alpha = 1-H$ have a similar effect as the existence of a non-trivial
quadratic variation.
 
\end{rem}

\section{Approximations}\label{sect:appr}

\subsection*{Binary tree approximations}

The famous Donsker's invariance principle links random walks to the Bm.
By using this principle one can approximate the BS model with 
Cox--Ross--Rubinstein (CRR) binomial trees.  To be more precise, let for
all $n\in\N$, $(\xi_k^n)_{k\in\N}$ be i.i.d. random variables with 
$\P[\xi_k^n=1]=1/2=\P[\xi_k^n=-1]$. Set
$$
W^n_t = \frac{1}{\sqrt{n}}\sum_{k=1}^{\lfloor nt\rfloor} \xi_k^n.
$$
Then the Donsker's invariance principle states that the processes $W^n$,
$n\in\N$, converge in the Skorohod space $D([0,T])$ to the Bm. Let $S^n$
to be the binomial model defined by
$$
S_t^n = \prod_{s\le t} \left(1+\Delta W_s^n\right).
$$
Then the processes $S^n$,$n\in\N$, converge weakly in 
$D([0,T])$ to the geometric Bm $S_t=e^{W_t-t/2}$, i.e. the binomial 
models $S^n$, $n\in\N$, approximate the BS model.

In \cite{Sottinen01} a fractional CRR model was
constructed that approximates the fBS model when $H>1/2$, and later this 
approximation was extended in different directions by Nieminen 
\cite{Nieminen04} and Mishura and Rode \cite{MishuraRode07}. We give 
here a brief overview of the construction in \cite{Sottinen01}:

Let $(\xi_k^n)_{k\in\N}$ be as before, and let $k(t,s)$ be the
kernel that transforms the Bm into a fBm:
$$
k(t,s) = c_H s^{\frac12-H}\int_s^t u^{H-\frac12}(u-s)^{H-\frac32}\, \d u,
$$
where
$$
c_H = (H-\frac12)\sqrt{\frac{(2H+\frac12)\Gamma(\frac12-H)}{
\Gamma(H+\frac12)\Gamma(2-2H)}},
$$
and $\Gamma(z)=\int_0^\infty t^{z-1}e^{-t}\d t$ is the Gamma function.
Then
$$
B_t = \int_0^t k(t,s)\, \d W_s.
$$
To get a piece-wise constant process in $D([0,T])$ one
must regularize the kernel:
$$
k^n(t,s) = n\int_{s-1/n}^s 
k\left(\frac{\lfloor nt \rfloor}{n},u\right)\, \d u.
$$
Set
$$
B^n_t = \int_0^t k^n(t,s)\,\d W^n_s
$$
and
$$
S_t^n = \prod_{s\le t} \left(1+\Delta B_s^n\right).
$$

\begin{thm}\label{thm:fCRR}
Let $H>1/2$.
\begin{enumerate}
\item
The random walks $B^n$, $n\in\N$, converge weakly in $D([0,T])$
to the fBm $B$.
\item
The binary models $S^n$, $n\in\N$, converge weakly in $D([0,T])$
to the fBS model $S=e^{B}$.
\item
The fractional CRR binary models $S^n$,$n\in\N$,
are complete, but exhibit arbitrage opportunities if $n$ is sufficiently large.
\end{enumerate}
\end{thm}

\begin{proof}
(i) is the ``fractional invariance principle'' 
\cite[Theorem 1]{Sottinen01}, (ii) follows basically from (i), the 
continuous mapping theorem, and a Taylor expansion of $\log(S^n)$, cf. the proof of 
\cite[Theorem 3]{Sottinen01} for details.  The completeness claim of 
(iii) is obvious, since the market models are binary. The arbitrage claim 
of (iii) follows from the fact that if we have only gone up in the 
binary tree for long enough, the stock-price will increase in the next 
step no matter which branch the process takes in the tree. We refer to 
the proof of \cite[Theorem 5]{Sottinen01} for details.
\end{proof}

A main motivation for considering the approximation $S^n$ is that the continuous time process $S_t=e^{B_t}$ solves the SDE
$$
dS_t=S_t dB_t,\quad S_0=1
$$
in the sense of forward integration. Alternatively, one can build an integral on Wick-Riemann sums \cite{Bender03, BiaginiHuOksendalZhang08, Duncan00, Mishura08} and examine the SDE
$$
dX_t=X_t d^\diamond B_t,\quad X_0=1.
$$
Here, $X_t=\exp\{B_t-t^{2H}/2\}$. Thus, the processes $S$ and $X$ only differ by a deterministic factor. Without going into any details here, we note that the \emph{Wick product} can be defined by
$$
e^{\Phi-E[\Phi^2]/2} \;\diamond \; e^{\Psi-E[\Psi^2]/2} = e^{(\Phi+\Psi)-E[(\Phi+\Psi)^2]/2}
$$
for centered Gaussian random variables $\Phi$ and $\Psi$ and can be extended to larger classes of random variables by bilinearity and denseness arguments, see e.g. \cite{Bender03, Duncan00}. Somewhat surprisingly, there is a very simple analogue of the Wick product for the binary random variables $\xi^n_k$, $k=1,\ldots, n$, see \cite{Holden92}, which gives rise to a natural binary discretization of $X_t$ suggested by Bender and Elliott \cite{BenderElliott04}.

The \emph{discrete Wick product} can be defined as ($A,B\subset \{1,\ldots,n\}$)
$$
\prod\limits_{i \in A} \xi^{n}_{i} \diamond_{n} \prod\limits_{i \in B} \xi^{n}_{i} \ := \ \left\{\begin{array}{cl} \prod\limits_{i \in A \cup B} \xi^{n}_{i} &\textnormal{if} \ A \cap B = \emptyset\\
0 &\textnormal{otherwise}
\end{array}\right.,$$
and extends by bilinearity to $L^2(\mathcal{F}_n)$, where $\mathcal{F}_n$ denotes the $\sigma$-field generated by $(\xi^n_1,\ldots,\xi_n^n)$. A discrete version of the Wick-fractional Black-Scholes model is then defined by
$$
X_t^n = \wick_{s\le t} \left(1+\Delta B_s^n\right).
$$
Bender and Elliott \cite{BenderElliott04} argue in favor of this discretization that the discrete Wick product separates influences of the drift and volatility. 

\begin{thm}\label{thm:WfCRR}
Let $H>1/2$.
\begin{enumerate}
\item
The binary models $X^n$, $n\in\N$, converge weakly in $D([0,T])$
to the Wick-fractional Black-Scholes model $X$.
\item
The Wick-fractional CRR binary models $X^n$,$n\in\N$,
are complete, but exhibit arbitrage opportunities if $n$ is sufficiently large.
\end{enumerate}
\end{thm}
The proof of (ii) is similar to the one of Theorem \ref{thm:fCRR}, (iii), and can be found in \cite{BenderElliott04}. As is pointed out there, the use of the discrete Wick products kills a part of the memory as compared to the discrete-time model $S^n$. It turns out, however, that the remaining part of the memory is still sufficient to construct an arbitrage. Completeness again follows from the fact that the model is binary.  For the proof of (i), one cannot argue by the continuous mapping theorem, because the discrete Wick product is not a pointwise operation. Instead the relation of the Wick powers to Hermite polynomials and explicit computations of the Walsh decomposition (which can be considered a discrete analogue of the chaos decomposition to some extent) can be exploited, see \cite{BenderParczweski09}.

\subsection*{Arbitrage-free approximation}
The results in this section are motivated by \cite{KluppelbergKuhn04}, where the authors give an arbitrage-free approximation to fBS model.  The prelimit models in this approximation are not complete, however.

Recall the following classical result: Let $N=(N_t)_{t\in\R_+}$ be a 
Poisson process with intensity $1$, and set
$$
W^n_t = \frac{1}{\sqrt{n}}\left(N_{nt}-nt\right).
$$
Then $W^n$ converges to a Bm $W$ in the Skorohod space $D([0,T])$, the process $dS^n _t = S^n_{t-}dW^n_t $, $S^n_0 =S_0$ converges weakly to the BS model, and the approximation is complete and arbitrage-free.  

We approximate the fractional Black--Scholes model $(S,( \F_t)_{t\in [0,T]} , \P )$  with a sequence 
$(S^n, (\F^n) _{t\in [0,T]}) $ of models driven by scaled renewal counting processes. The prelimit models are complete and arbitrage-free.
The approximation is based on the limit theorem of Gaigalas and Kaj
\cite{GaigalasKaj03}. 
It goes as follows: let $G$ be a continuous distribution function with heavy tails. i.e.  
\begin{equation}\label{eq:h-tails}
1-G(t) \sim t^{-(1+\beta )} 
\end{equation}
as $t\to \infty$ with $\beta \in (0,1)$.

Take  $\eta _i$ to be  the sojourn times of a renewal counting process $N$. Assume that 
$\eta _i \sim G $ for $i\ge 2$; for the first sojourn time $\eta _1$ assume that it has the 
distribution $G_0(t)= \frac{1}{\mu} \int _0^t (1-G(s))ds$ (here $\mu$ is the normalizing constant), so that the renewal counting process 
$$
N_t =  \sum _{k=1}^\infty 1_ {\{\tau _k \le t \}}
$$
is stationary, where $\tau _1 = \eta _ 1$ and $\tau _k := \eta _1 + \cdots + \eta _k $.

Take now independent copies $N^{(i)}$ of $N$, numbers $a_m\ge 0$, $a_m\to \infty$ such that 
\begin{equation}\label{eq:htl-1}
\frac{m}{a_m^\beta } \to \infty ;
\end{equation}
using the terminology of Gaigalas and Kaj we can speak of \emph{fast connection rate}.

Define the \emph{workload}
process $W(m,t)$ by 
$$
W(m,t) = \sum _{i=1}^m N^{(i)}_t ;
$$ 
note that the process $N^m$ is a counting process, since the sojourn distribution is continuous.
We have that $E W(m,t) = \frac{mt}{\mu}$, since $W(m,t)$ is a stationary process.

\begin{pro}[Gaigalas and Kaj \cite{GaigalasKaj03}] \ \ Assume \eqref{eq:h-tails} and \eqref{eq:htl-1}. Let 
$$
Y ^m (t) := \mu ^{\frac32}\sqrt{\frac{\beta(1-\beta)(2-\beta)}{2}}\frac{W(m,a_mt) - m\mu^{-1}a_mt }{m^{\frac12 }
a_m^{1-\frac{\beta}{2}} } .
$$
Then $Y^m$ converges weakly [in the Skorohod space $D$] to a fractional Brownian motion $B^H$, where 
$H= 1-\frac{\beta}{2}$.
\end{pro}


Since the process 
$Y^m$ is a semimartingale, it has a  semimartingale decomposition 
\begin{equation}\label{eq:htl-2}
Y^m = M^m + H^m; 
\end{equation}
here $H^m = B^m -A^m $ and $B^m$ is the compensator of the normalized aggregated counting process 
$W$. Note that the process $H^m$ is a continuous process with bounded variation.

Up to a constant we have that 
the square bracket of the martingale part $M^m$ of the semimartingale $Y^m$ is 
$$
[M^m,M^m]_t = C \frac{W(m,a_mt)}{ma^{2-\beta} _m}.
$$

But our assumption imply that $[M^m,M^m]_t\stackrel{L^1(P)}{\to}0$, as $m\to \infty$. With the 
Doob inequality we obtain that $\sup _{s\le t} | M^m_s| \stackrel {P}{\to} 0$, and fBm is the limit
of a sequence of continuous processes with bounded variation. 

It is not difficult to check that the solution to the linear equation 
$$
dS^m_t = S^m_{t-} \d Y^m_s 
$$
converges weakly in the Skorohod space to geometric fractional Brownian 
motion.

The driving process $Y^m$ is a scaled counting process minus the 
expectation. It is well known that such models are complete and 
arbitrage-free. Hence we have a complete and arbitrage-free 
approximation to fractional Black--Scholes model. See \cite{Valkeila09} 
for more details.

\begin{rem}
If one computes the hedging price and the hedging strategy for the 
European call $(S^m_T -K)^+ $ in  the prelimit sequence, and lets 
$m\to \infty$ one gets in the limit the stop-loss-start-gain hedging 
given in Example \ref{ex:fbs-arbitrage}.
\end{rem}

%

\subsection*{Microeconomic approximation}

So far there has been few economic justifications to use fractional
models in option-pricing. E.g the LRD of the stock-price, measured by the 
Hurst index $H$, is usually given as an econometric fact (and even that
is questionable). One attempt to build a microeconomic foundation for 
fractional models was that of Bayraktar et al. 
\cite{BayraktarHorstSircar06}. They showed how the fBS model 
can arise as a large time-scale many-agent limit when there are inert 
agents, i.e. investors who change their portfolios infrequently,
and the log-price is given by the market imbalance.  We 
will briefly explain their framework and their main result here.

Consider $n$ agents. Each agent $k$ has a \emph{trading mood}
$x^k=(x^k_t)_{t\in[0,\infty)}$ that takes values in a finite state-space
$E\subset\R$ containing zero: $x^k_t>0$ means buying, $x^k_t<0$ means 
selling, and $x_t^k=0$ means inactivity at time $t$.
The agents are homogeneous and independent. The trading mood $x^k$ is a 
semi-Markov process defined as
$$
x_t^k = \sum_{m=0}^\infty 
\xi_m^k \1_{\big[\tau_m^k,\tau_{m+1}^k\big)}(t),
$$
where the $E$-valued random variables $\xi_m^k$ and the stopping times
$\tau_m^k$ satisfy 
\begin{eqnarray*}
\lefteqn{\P\left[\xi_{m+1}^k=j,\tau_{m+1}^k-\tau_m^k\le t \,\bigg|\,
\xi_1^k,\ldots,\xi_m^k,\tau_1^k,\ldots,\tau_m^k\right]} \\
&=&
\P\left[\xi_{m+1}^k=j, \tau_{m+1}^k-\tau_m^k \le t
\,\bigg|\, \xi_m^k\right] \\
&=& Q(\xi_m^k,j,t).
\end{eqnarray*}
So, $(\xi_m^k)_{m\in\N}$ is a homogeneous Markov chain on $E$ with
transition probabilities 
$p_{ij}=\lim_{t\to\infty} Q(i,j,t)$. It is assumed that $p_{ij}>0$
for all $i\ne j$ so that $(p_{ij})$ admits a unique stationary measure
$\P^*$. On the sojourn times $\tau_{m+1}^k-\tau_m^k$ given 
$\xi_{m}^k$ it is assumed that:
\begin{enumerate}
\item The average sojourn times are finite.
\item The sojourn time at the inactive state is heavy-tailed, i.e. there 
exist a constant $\alpha\in(1,2)$ and a locally bounded slowly varying at 
infinity function $L$ such that
$$
\P\left[\tau_{m+1}^k-\tau_m^k \ge t \,\big|\,\xi_m^k=0\right] 
\sim t^{-\alpha}L(t).
$$
($L$ is slowly varying at infinity if, for all $x>0$, $L(xt)/L(t)\to 1$,
when $t\to\infty$).
\item The sojourn times at the active states $i\ne 0$ are 
lighter-tailed than the sojourn time at the inactive state:
$$
\lim_{t\to\infty} \frac{\P[\tau_{m+1}^k-\tau_m^k\ge t|\xi_m^k=i]}{
t^{-(\alpha+1)}L(t)} =0.
$$
\item
The distribution of the sojourn times have continuous and bounded
densities with respect to the Lebesgue measure.
\end{enumerate}

An agent-independent process $\Psi=(\Psi_t)_{t\in[0,\infty)}$ describes 
the sizes of typical trades: Agent $k$ accumulates the asset $S$ at the 
rate $\Psi_t x^k_t$ at time $t$. The process $\Psi$ is assumed to be a 
continuous semimartingale with Doob--Meyer decomposition $\Psi=M+A$ such 
that $\E[\la M\ra_T]<\infty$ and  $\E[\mathcal V(A)]<\infty$, and 
$\Psi$ and the $x^k$'s are independent. As before, $\mathcal V(A)$ denotes the total variation of $A$ on $[0,T]$.

The log-price $X^n$ for the asset with $n$ agents is assumed to be
given by the \emph{market imbalance}:
$$
X_t^n = X_0 + \sum_{k=1}^n \int_0^t \Psi_s x_s^k\, \d s.
$$

The \emph{aggregate order rate} is
$$
Y^{\eps,n}_t = \sum_{k=1}^n \Psi_t x^k_{t/\eps}.
$$
Let $\mu\ne 0$ be the expected trading mood under the stationary measure
$\P^*$, and define the centered aggregate order process
$$
X_t^{\eps,n} = \int_0^t Y_s^{\eps,n} \, \d s - 
\mu n \int_0^t \Psi_s\, \d s.
$$
Then, the main result \cite[Theorem 2.1]{BayraktarHorstSircar06} states
that in the limit the centered log-prices are given by a stochastic 
integral with respect to a fBm:

\begin{thm}\label{thm:microlimit}
There exists a constant $c>0$ such that
$$
\lim_{\eps\downarrow 0}\lim_{n\to\infty}
\frac{1}{\eps^{1-H}\sqrt{n L(1/\eps)}}X^{\eps,n}
= c\int_0^\cdot \Psi_t\, \d B_t,
$$
where $B$ is a fBm with Hurst index $H=(3-\alpha)/2 > 1/2$.
The limits are weak limits in the Skorohod space $D([0,T])$. 
\end{thm}

\begin{rem}
Assume that $\Psi\equiv 1$, i.e. the trades, and consequently the 
log-prices, are completely determined by the agents' intrinsic trading
moods. Then
$$
X^{\eps,n} = \eps X^n_{t/\eps} -\mu n t.
$$
\begin{enumerate}
\item
The limit in Theorem \ref{thm:microlimit} is the fBS model.
\item
Bayraktar et al. \cite{BayraktarHorstSircar06} also considered a model 
where there are both active and inert investors 
(active investors have  light-tailed sojourn times at the inactive state 
$0$). Then they get, in the limit, the mfBS model.
\end{enumerate}
\end{rem}

\section{Conclusions}

We have given some recent results on the arbitrage and hedging in some fractional pricing models. 
If one wants to understand the pricing of options in the fBS model, then it is not clear to what extent the hedging capital given in \eqref{eq:fbs-hedge} can be interpreted as the price of the option.  On the other hand, these exact hedging results may have some value, if one studies the hedging problem in the presence of transaction costs. The mixed Brownian-fractional Brownian pricing model has less arbitrage possibilities, but it is possible to model the  dependency of the log-returns in this model family. One can also modify this model to include more 'stylized' properties of log-returns, but the hedging prices will be the same as without these 'stylized' features.

The mixed model seems to be a good candidate to include several of the observed 'stylized' facts of log-returns in the modelling of stock prices. Hence it is reasonable to study how the properties of the standard gBS model change in the mixed model. We have shown in \cite{BenderSottinenValkeila08} that the hedging is the same in all models having the same structural quadratic variation as a functional of the stock price path. For example, recently
Bratyk and Mishura have considered quantile hedging problems in mixed models; see  \cite{BratykMishura08} 
for more details.

\subsection*{Open problems}
We finish by giving some open problems related to the present survey.

Are fractional and mixed models free of simple arbitrage?

What kind of random variables have a Riemann-Stieltjes integral representations in the fBS model?

Can one verify statistically that option prices depend only on the quadratic variation of the underlying 
stock prices?

What is the best way to estimate quadratic variation?


\begin{thebibliography}{10}
\bibitem{AndroshcukMishura06}
\textsc{Androshchuk, T. and Mishura, Yu.}
Mixed Brownian--fractional Brownian model: absence of arbitrage and 
related topics.
\emph{Stochastics}  \textbf{78}  (2006),  no. 5, 281--300.

\bibitem{Arcones95}
\textsc{Arcones, M. A.} 
On the law of the iterated logarithm for Gaussian processes.  
\emph{J. Theoret. Probab.}  \textbf{8}  (1995),  no. 4, 877--903.

\bibitem{Azmoodeh09}
\textsc{Azmoodeh, E.} 
Geometric fractional Brownian motion market model with transaction costs.
\emph{Preprint}, (2009), 12 p.  

\bibitem{AzmoodehMishuraValkeila09}
\textsc{Azmoodeh, E., Mishura, Yu., and Valkeila, E.}
On European options in geometric fractional Brownian motion market model.
\emph{Statist. Decisions}, (2010), forthcoming.

\bibitem{BayraktarHorstSircar06}
\textsc{Bayraktar, E., Horst, U., and Sircar, R.}
A limit theorem for financial markets with inert investors.  
\emph{Math. Oper. Res.}  \textbf{31}  (2006),  no. 4, 789--810. 

\bibitem{Bender03}
\textsc{Bender, C.}
An $S$-transform approach to integration with respect to a fractional Brownian motion. 
\emph{Bernoulli} \textbf{9} (2003), no. 6, 955-983.

\bibitem{BenderElliott04}
\textsc{Bender, C. and Elliott, R. J.} 
Arbitrage in a discrete version of the Wick-fractional Black-Scholes 
market. 
\emph{Math. Oper. Res.}  \textbf{29}  (2004),  no. 4, 935--945. 

\bibitem{BenderParczweski09}
\textsc{Bender, C. and Parczewski, P.} 
Approximating a geometric fractional Brownian motion and related processes via discrete Wick calculus.
\emph{Bernoulli} (2009), to appear.



\bibitem{BenderSottinenValkeila08}
\textsc{Bender, C., Sottinen, T., and Valkeila, E.}
Pricing by hedging and no-arbitrage beyond semimartingales.  
\emph{Finance Stoch.}  \textbf{12}  (2008),  no. 4, 441--468.

\bibitem{BiaginiHuOksendalZhang08}
\textsc{Biagini, F., Hu, Y., {\O}ksendal, B., and Zhang, T.}
\emph{Stochastic calculus for fractional Brownian motion and 
applications.} Probability and its Applications (New York). 
Springer-Verlag London, Ltd., London, 2008.

\bibitem{BjorkHult05}
\textsc{Bj\"{o}rk, T. and Hult, H.}
A note on Wick products and the fractional Black-Scholes model.  
\emph{Finance Stoch.}  \textbf{9}  (2005),  no. 2, 197--209. 

\bibitem{BratykMishura08}
\textsc{Bratyk, M., and Mishura, Y.} 
The generalization of the quantile hedging problem for price process model involving finite number of Brownian and fractional Brownian motions.
\emph{Theory Stoch. Process.}  \textbf{14}  (2008),  no. 3-4, 27--38.

\bibitem{Cheridito01}
\textsc{Cheridito, P.}
Mixed fractional Brownian motion. 
\emph{Bernoulli}  \textbf{7}  (2001),  no. 6, 913--934. 

\bibitem{Cheridito03}
\textsc{Cheridito, P.} 
Arbitrage in fractional Brownian motion models.  
\emph{Finance Stoch.}  \textbf{7}  (2003),  no. 4, 533--553.

\bibitem{Cherny08}
\textsc{Cherny, A.} 
Brownian moving averages have conditional full support.
\emph{Ann. Appl. Probab.}  \textbf{18}  (2008),  no. 5, 1825--1830.

\bibitem{DasguptaKallianpur00}
\textsc{Dasgupta, A. and Kallianpur, G.} 
Arbitrage opportunities for a class of Gladyshev processes.  
\emph{Appl. Math. Optim.}  \textbf{41}  (2000),  no. 3, 377--385. 

\bibitem{DelbaenSchachermayer94}
\textsc{Delbaen, F. and Schachermayer, W.}
A general version of the fundamental theorem of asset pricing.  
\emph{Math. Ann.}  \textbf{300}  (1994),  no. 3, 463--520.

\bibitem{DelbaenSchachermayer98}
\textsc{Delbaen, F. and Schachermayer, W.} 
The fundamental theorem of asset pricing for unbounded stochastic
processes.  
\emph{Math. Ann.}  \textbf{312}  (1998),  no. 2, 215--250.

\bibitem{Duncan00}
\textsc{Duncan, T. E., Hu, Y. and Pasik-Duncan, B.}
Stochastic calculus for fractional Brownian motion. I: Theory.
{\it SIAM J. Control Optimization} {\bf 38}, no. 2, (2000), 582--612.



\bibitem{Foellmer81}
\textsc{F{\"o}llmer, H.} 
 Calcul d'It\^o sans probabilit\'es.
\emph{S\'eminaire de Probabilit\'es, XV}, 143--150, Springer,
Berlin (1981).

\bibitem{GaigalasKaj03}
\textsc{Gaigalas, R. and Kaj, I.}
Convergence of scaled renewal processes and a packet arrival model.  
\emph{Bernoulli}  \textbf{9}  (2003),  no. 4, 671--703.



\bibitem{GasbarraSottinenVanZanten08}
\textsc{Gasbarra, D., Sottinen, T., and van Zanten, H.} 
Conditional full support of Gaussian processes with stationary 
increments. \emph{Preprint}, 2008. 

\bibitem{Guasoni06}
\textsc{Guasoni, P.}
No arbitrage under transaction costs, with fractional Brownian motion
and beyond.  
\emph{Math. Finance}  \textbf{16}  (2006),  no. 3, 569--582.


\bibitem{GuasoniRasonyiSchachermayer08a}
\textsc{Guasoni, P., R\'{a}sonyi, M., and Schachermayer, W.}
Consistent price systems and face-lifting pricing under transaction
costs.  
\emph{Ann. Appl. Probab.}  \textbf{18}  (2008),  no. 2, 491--520.

\bibitem{GuasoniRasonyiSchachermayer08}
\textsc{Guasoni, P., R\'{a}sonyi, M., and Schachermayer, W.}
The fundamental theorem of asset pricing for continuous processes under small transaction costs. 
\emph{Ann. Finance}  \textbf{6}  (2010),  no. 2,  157--191.


\bibitem{Holden92} 
\textsc{Holden, H., Lindstr{\o}m, T., {\O}ksendal, B. and Ub{\o}e, J.} 
Discrete Wick calculus and stochastic functional equations.
{\it Potential Anal.} {\bf 1} (1992), no. 3, 291--306.


\bibitem{JarrowProtterSayit09}
\textsc{Jarrow, R., Protter, P., and Sayit, H.} 
No Arbitrage without Semimartingales, 
\emph{Ann. Appl. Probab.} {\bf 19} (2009), no. 2, 596--616. 

\bibitem{JouiniKallal95}
\textsc{Jouini, E., and Kallal, H.} 
Martingales and arbitrage in securities markets with transaction costs.
\emph{J. Econom. Theory}, {\bf 66} (1995), 178--197.
\bibitem{KarlinTaylor75}
\textsc{Karlin, S. and Taylor, H. M.}
\emph{A first course in stochastic processes.} Second edition. 
Academic Press, New York-London, 1975.

\bibitem{KluppelbergKuhn04}
\textsc{Kl\"{u}ppelberg, C. and K\"{u}hn, C.}
Fractional Brownian motion as a weak limit of Poisson shot noise
processes---with applications to finance.  
\emph{Stochastic Process. Appl.} \textbf{113} (2004),  no. 2, 333--351.

\bibitem{Mishura08}
\textsc{Mishura, Yu. S.} 
\emph{Stochastic calculus for fractional Brownian motion and 
related processes.} Lecture Notes in Mathematics, \textbf{1929}. 
Springer-Verlag, Berlin, 2008.

\bibitem{MishuraRode07}
\textsc{Mishura, Yu. S. and Rode, S. G.} 
Weak convergence of integral functionals of random walks that weakly 
converge to fractional Brownian motion. (Ukrainian)  
\emph{Ukrain. Mat. Zh.}  \textbf{59}  (2007),  no. 8, 
1040--1046;  translation in
\emph{Ukrainian Math. J.}  \textbf{59}  (2007),  no. 8, 1155--1162


\bibitem{Nieminen04}
\textsc{Nieminen, A.}
Fractional Brownian motion and martingale-differences.  
\emph{Statist. Probab. Lett.}  \textbf{70}  (2004),  no. 1, 1--10.


\bibitem{Pakkanen09}
\textsc{Pakkanen, M. S.} 
Stochastic integrals and conditional full support. 
\emph{Preprint.} 
(2008, Revised Jul. 24, 2009) [Available as arXiv:0811.1847]

\bibitem{Protter04}
\textsc{Protter, P.}
\emph{Stochastic integration and differential equations.} 
Second edition. Applications of Mathematics (New York), \textbf{21}. 
Stochastic Modelling and Applied Probability. Springer-Verlag, Berlin, 
2004.


\bibitem{Rogers97}
\textsc{Rogers, L. C. G.} 
Arbitrage with fractional Brownian motion.  
\emph{Math. Finance}  \textbf{7}  (1997),  no. 1, 95--105.


\bibitem{RussoVallois93}
\textsc{Russo, F. and Vallois, P.} 
 Forward, backward and symmetric
stochastic integration. 
\emph{Probab. Theory Related Fields } {\bf
97} (1993), no. 3, 403--421.

\bibitem{SchoenmakersKloeden99}
\textsc{Schoenmakers, J. G. M. and Kloeden, P. E.} 
Robust option replication for a Black-Scholes model extended with
nondeterministic trends.  
\emph{J. Appl. Math. Stochastic Anal.} \textbf{12}  (1999),  no. 2, 
113--120.

\bibitem{Shiryaev98}
\textsc{Shiryaev, A.}
On arbitrage and replication for fractal models.
\emph{Research report} \textbf{30}, (1998),
MaPhySto, Department of Mathematical Sciences, University of Aarhus.

\bibitem{Sottinen01}
\textsc{Sottinen, T.} 
Fractional Brownian motion, random walks and binary market models.  
\emph{Finance Stoch.}  \textbf{5}  (2001),  no. 3, 343--355. 

\bibitem{SottinenValkeila03}
\textsc{Sottinen, T. and Valkeila, E.}
On arbitrage and replication in the fractional Black-Scholes pricing 
model.  
\emph{Statist. Decisions}  \textbf{21}  (2003),  no. 2, 93--107.

\bibitem{Valkeila09}
\textsc{Valkeila, E.} 
On the approximation of geometric fractional Brownian motion.
In: \emph{Optimality and Risk - Modern Trends in Mathematical Finance
The Kabanov Festschrift.}
Delbaen, F., R{\'a}sonyi, M., and  Stricker, C. (Eds.) (2009), 251--265. 
\end{thebibliography}
\end{document}